\documentclass[pre,aps,twocolumn]{revtex4}
\usepackage{amsmath,bm,amssymb}
\usepackage[english,russian]{babel}
\usepackage{color}

\newcounter{prob}

\newcommand{\const}{\operatorname{const}}

\newcommand{\Ku}{\operatorname{Ku}}

\newcommand{\diverg}{\operatorname{div}}
\newcommand{\St}{\operatorname{St}}
\begin{document}










\title{Localization-delocalization transitions in turbophoresis of inertial particles
} 
\author{Sergei Belan$^{1,2}$, Itzhak Fouxon$^3$ and Gregory Falkovich$^{3,4}$}
\affiliation{$^1$Moscow Institute of Physics and Technology, Dolgoprudny, Russia \\$^2$Landau Institute for Theoretical Physics, Moscow, Russia\\$^3$Weizmann Institute of Science, Rehovot,  Israel \\$^4$Institute for Information Transmission Problems, Moscow, Russia}


\begin{abstract} Small aerosols drift down temperature or turbulence gradient since faster particles fly longer distances before equilibration. That fundamental phenomenon, called thermophoresis or turbophoresis, is  widely encountered in nature and used in industry \cite{Kampen,Milligen,Widder_1989,Stolovitzky_1998,lopez,Reeks}. It is universally believed that particles moving down the kinetic energy gradient must concentrate in minima (say, on walls in turbulence). Here we show that this is incorrect: escaping minima is possible for inertial particles whose time of equilibration is longer than the time to reach the minimum.
The best way out is always through: particles escape by flying through minima or reflecting from walls. We solve the problem analytically and find the phase transition as a sign change of the mean velocity.  That means separation: light particles concentrate in a minimum while heavy particles spread away from it (gravity can reverse the effect). That discovery changes understanding of that fundamental phenomenon and may find numerous applications.
\end{abstract}\maketitle
Small particles in a fluid are agitated by thermal or turbulent fluctuations. In a spatial gradient of temperature or turbulence intensity, an inertial particle which flies from the hot side moves faster and travels longer distance before equilibration than a particle flying from a colder side. That produces net flux of particles towards the minima of kinetic energy of the fluid. Discovery of thermophoresis, wherein particles go with the heat flow,  goes back to Maxwell, Tyndall and Rayleigh. A common example is the blackening of the kerosene lantern glass: the temperature gradient drives the carbon particles away from the flame. There are special cases where the sign of a drift can be opposite (like for particles driven by a colored noise \cite{Scott}, some cases of Soret effect in liquids etc) but we consider here the simplest situation.
Since particles flow towards temperature minima, it was invariably concluded that they must accumulate there. That statement was then generalized from thermal diffusion to turbulent diffusion predicting the so-called turbophoresis: it is widely accepted that heavy particles  migrate toward the turbulence minima, particularly to walls  \cite{Reeks,Ci,Reek,McL,Brook}. Turbophoresis is ubiquitous in nature and industry, from droplets in clouds to coal fire burners; it determines, for instance, how chemicals remaining near walls are washed out into the bulk of reactor,  how dust is raised from the ground by a turbulent wind or deposited on surfaces. Thermophoresis also is of interest both as a fundamental phenomenon and because of practical importance in many industrial applications, such as thermal precipitators removing sub-micron sized particles from gas streams,  laser Doppler velocimetry, manufacturing by vapor deposition, manipulations of nano-particles, etc \cite{Talbot}.

Here we predict a new phenomenon: sufficiently inertial particles can have a net flux against the gradient, escaping kinetic energy minima. We show below that upon the change of the inertia parameter, the system undergoes localization-delocalization transition. We consider idealized models which allow analytic description and illuminate clearly the nature of the phenomenon. Yet our models are rich enough to account for other characteristics  having influence on this transition: correlation time of the flow, external force and boundary conditions on the wall.

Consider small inertial particle in a flow, whose mean intensity of fluctuations depends on $z$ coordinate. Generally, there are three characteristic times in the problem: $\tau_c$ is the fluid flow correlation time {\em in the reference frame of the particle}, $\tilde\tau(z)$ is the time it takes for a particle flying with a mean velocity to feel flow inhomogeneity and $\tau=1/\gamma=2a^2\rho_0/9\nu\rho$ is the particle relaxation (Stokes) time expressed via the particle radius $a$ and density $\rho_0$, and fluid viscosity $\nu$ and density $\rho$. The equations on position $z$ and velocity $v$ have the form $\dot z=v$, $\dot v=\gamma (u-v)$, assuming Stokes drag.
We start from the most common analytic approach based on assuming  ${\tau_c}$ being the shortest time, considering more general case at the end. Taking the fluid velocity random Gaussian with $\langle u(z,0)u(z,t)\rangle=2\kappa(z)\delta(t)$, one obtains the standard Fokker-Planck equation for the probability density function (pdf) over positions and velocities, $\rho(z,v)$, see e.g. \cite{Kampen,Milligen,Widder_1989,Stolovitzky_1998,lopez,Reeks}:
\begin{equation}
\partial \rho/\partial t=-v \partial_z \rho+\gamma\partial_{v }(v \rho)+\gamma^2\kappa(z)\partial_{v }^2\rho\ .\label{FP1}
\end{equation}
In this limit, the medium inhomogeneity is characterized by the height-depending effective temperature of noise $\kappa(z)$, which determines the diffusivity in the velocity space $\gamma^2\kappa$. While thermal fluctuations can always be considered short-correlated, turbulent fluctuations can be treated as such only for sufficiently inertial particles: (\ref{FP1}) is valid when  the Stokes number $\St=\tau/\tau_c$ is large. We here focus on the measure of inertia relative to inhomogeneity,  $\tau/\tilde\tau(z)$, which can be both large and small depending on the inhomogeneity. As far as inhomogeneous temperature is concerned, the friction coefficient $\gamma$ depends on viscosity, which generally depends on the temperature. We, however, use here the common approximation of space-independent $\gamma$, leaving  straightforward generalizations for future work.

Now we focus our attention on the motion near minimum of noise intensity determined by $\kappa(z)$.
Let us assume that the motion takes place in the full one-dimensional domain and $\kappa(z)=\mu z^2$, where $\mu$ is the measure of fluid velocity fluctuations intensity.
To get the respective characteristic time  $\tilde\tau(z)$, we first estimate the typical velocity by comparing the second and last terms in the rhs of (\ref{FP1}), $\bar v(z)\simeq (\gamma \kappa)^{1/2}$  and then divide the scale of change $\kappa/\kappa'$ by it: $\tilde\tau(z)\simeq(\mu \gamma)^{-1/2}$,
that is  the degree of inertia $I=(\mu\tau)^{1/3}=[\tau/\tilde\tau(z)]^{2/3}$ is $z$-independent.

While one cannot describe analytically the evolution of an arbitrary initial distribution even for the quadratic case, there is an elegant way to show the direction of evolution and the related transition.
Let us introduce the variable
$\sigma=v/z$, which obeys a closed equation
\begin{eqnarray}
\dot{\sigma}=\gamma(s-\sigma)-\sigma^2=\gamma s-\partial U/\partial \sigma \label{sigma}
\end{eqnarray}
with $\langle s(0)s(t)\rangle=2\mu\delta(t)$ and $U(\sigma)= {\gamma\sigma^2}/{2}+ {\sigma^3}/{3}$. The probability distribution $P(\sigma,t)$ satisfies the equation \begin{equation}\partial_tP=\partial_{\sigma}\left[U'P\right]+\mu\gamma^2\partial_{\sigma}^2 P=0\ .\label{FPSigma}\end{equation}
There is extensive literature devoted to this equation,   it describes quantum particle  in a random potential $s$, two particles in random flows etc, see e.g. \cite{Halperin,Gaw2,Deutch, WM, FT}. Here we use few elementary properties of it. Gibbs state $P(\sigma)=\exp[-U(\sigma)/\mu\gamma^2]$ is not normalizable. However, $\sigma$-space has the topology of a circle with $\sigma=-\infty$ glued to $\sigma=+\infty$, which corresponds to a particle flying though the minimum from either side or reflecting off an elastic wall at $z=0$. That topology allows the flux of probability going towards $-\infty$ and returning from  $+\infty$ (the mean rate of flying through zero or reflecting from the wall): $U'P+\mu\gamma^2P'=F$. The normalizable steady state,
\begin{equation}P(\sigma)={F\over\mu\gamma^2}e^{-U(\sigma)/\mu\gamma^2}\int_{-\infty}^{\sigma}\!\!\!
e^{U(\sigma')/\mu\gamma^2} d\sigma'\,,\label{Psigma}\end{equation}    realizes the minimum of entropy production \cite{FT} and is the asymptotic state of any initial distribution \cite{Gaw2}.

While $P(\sigma)$ fast acquires the stationary form  (\ref{Psigma}),  the evolution in the whole phase space depends crucially on the parameter $ I$. The reason is that
the Lyapunov exponent  $\langle\sigma\rangle=\langle v /z\rangle= t^{-1}\lim_{t\to\infty}\langle \ln [z(t)/z(0)]\rangle$, which determines the long-time evolution of the particle position,  is negative at $I<I_c=0.827$ and positive otherwise \cite{Halperin, Deutch, WM}. Particle with weak inertia drift towards the minimum and concentrate there, particles with strong inertia drift away. While the flux in $\sigma$-space is equal to $-F$ and always negative, the flux in the real space, $\int_{-\infty}^{\infty}\rho vdv\propto sign(z)\langle\sigma\rangle$, has different signs for different inertia, see the Appendix.
This demonstrates the localization-delocalization phase transition: infinitesimal change of the control parameter $I$ changes the Lyapunov exponent and the flux of particles in the whole space. In other applications of (\ref{sigma},\ref{FPSigma}), this  is known as order-disorder transition \cite{Deutch} and path-coalescence transition \cite{WM}.

How this phase transition manifests itself depends on the initial and boundary conditions in the whole phase space. Flux and equilibrium solutions of (\ref{FP1}) are straightforward to find. The  solution with a constant flux along $z$ follows from  (\ref{Psigma}): $\rho(z,v )\propto z^{-2}P(v/z)$. It has the concentration $n(z)=\int \rho(z,v )\,dz\propto 1/z$ whose regularization we discuss below.
When there are neither sources nor sinks, one may expect to observe the equilibrium solution with a zero flux in $z$-space, which has a simple form $\rho(v,z)=\delta(v)\delta(z)$. However,  it is clear from the above consideration that such solution is established only for $\langle v /z\rangle<0$  i.e.  at $I<I_c$. For $I>I_c$, the equilibrium solution is unstable and is not an asymptotic state of evolution because of the wrong sign of the flux, there is no steady state and the particles escape.

Intuitively, the escape can be related to inertial particles flying through the minimum (or reflecting from a wall) with the rate $F$. Indeed, for weak inertia at  $\mu \tau\ll I_c$ the typical particle velocity is $\bar v \simeq  (\kappa /\tau)^{1/2}=(z/\tau)(\mu \tau)^{1/2}$ and $z/\bar v \simeq   \tau(\mu \tau)^{-1/2}\gg\tau$. That means that most particles equilibrate locally and the fluxes are determined by a small vicinity; particles coming from lower $z$ have lower velocity than those from above, so that the net flux is towards $z=0$. On the contrary, for strong inertia at  $\mu \tau\gg I_c$
the flight time from any $z$ to zero is much smaller than $\tau$: $z/\bar v \simeq   \tau(\mu \tau)^{-1/2}\ll\tau$. Now, even though particles coming from above any given $z$ are faster, most of them do not feel (weaker) turbulence below, ballistically fly through and escape via $-z$, so that particles coming from below $z$ provide for a net positive flux  (particles over-shooting minimum were observed in numerical simulations of mixing layers \cite{MM}). The flow unwinds the particle that on each reflection from the region of stronger turbulence at large $|z|$ penetrates to larger $|z|$ on the opposite side so that its velocity increases exponentially. 
Despite going away exponentially particles return to zero with the constant rate $F$ of finite-time blow ups of $\sigma$ that occur due to $\sigma^2$ term in Eq.~(\ref{sigma}). The time spent near the origin decreases exponentially.

Let us now show that weakly/strongly inertial particles always go down/up the gradient in a steady flux state for arbitrary monotonic dependence $\kappa(z)$. The limit of weak inertia corresponds to the overdamped regime in which the local equilibrium is established with the Maxwell distribution over velocities with local $\kappa(z)\tau$ playing the role of temperature.
Integrating over $v$ one can obtain a reduced description of the system so that the only variable explicitly remaining
is the position of the particle, see the Appendix.
Then the equation for the concentration on timescales exceeding small time $\tau$ is as follows:
\begin{equation}\label{kin1}
\partial_t n=\partial_z^2 [\kappa(z) n]\ .
\end{equation}
The equilibrium (fluxless) steady state,  $n(z)\propto 1/\kappa(z)$, has maxima in the minima of diffusivity, which is the usual thermo- and turbophoresis: weakly inertial particles flow along the diffusivity gradient and concentrate in the minima of diffusivity. The steady state with non-zero flux $J$ has the form  $n(z)=-J |z|/\kappa(z)$, which decays with $z$ slower than equilibrium, and the flux is negative (tending to restore equilibrium \cite{ZLF}).

Consider now the limit of strong inertia when the second term in the rhs of  (\ref{FP1}) can be neglected. Let us pass from $z$ to $w$ such that $dw/dz=\gamma^2\kappa(z)$ (which works for $z$-dependent $\gamma$ as well). The steady solution of the Fokker-Planck equation, $-v\partial_w \rho+  \partial_{v}^2\rho=0$, can be sought in the form $\rho(v,w)=w^{-b}h(\xi)$
where $\xi=v^3/w$. The equation then takes the form $9\xi^{1/3}\partial_\xi\xi^{2/3}h'=-\xi h'-bh$.
Integrating over $\xi$ we obtain (disregarding zero or infinite boundary terms)
\begin{equation}
(b-2/3)\int_{-\infty}^{+\infty} h\xi^{-1/3} d\xi\propto (b-2/3)\int_{-\infty}^{+\infty} \rho v dv=0\,.
\end{equation}
Therefore finite non-zero flux requires $b=2/3$, which makes the equation readily integrable giving
\begin{equation}
\rho(v,z)\propto w^{-2/3} e^{-v^3/9w }\int_{-\infty}^{v^3/w}e^{\xi/9 }\xi^{-2/3}d \xi \ .\label{Strong}
\end{equation}
$J=\int_{-\infty}^{+\infty}\rho vdv\propto\int_{-\infty}^{+\infty}dx \cdot x e^{-x^3}\int_{-\infty}^{x}e^{y^3}d y\approx 0.6$ - the flux is positive. We see that the flux has different signs for weak and strong inertia. Note that $n(z)\propto w^{-1/3}$.

For $\kappa(z)=\mu z^m$ with $m\not=2$, the characteristic time  $\tilde\tau(z)$ and the inertia parameter $\tau/\tilde\tau(z)$ are position dependent. The condition $\tilde\tau(z^*)=\tau$ gives the level $z^*= (\mu\tau)^{{1}/{(2-m)}}$. For $m$ larger/smaller than $2$ the inertia dominates over inhomogeneity at large/small $z$ comparing to $z^*$. For example, for a turbulent boundary layer with $\kappa(z)=v_*z$ \cite{MY,ZA} we obtain $z^*=v_*\tau$, so different particles can be localized at different distances from the wall. For dust particles in the air we have $\rho_0/\rho\simeq10^{3}$ and $\nu=0.15\,cm^2/sec$  so that the Stokes time in seconds is  $\tau=1.3\cdot10^{3}a^2$, where the particle radius $a$ is in centimeters. For $a=10^{-2}\,cm$ and a moderate wind with $v_*\simeq 10\,m/sec$ we estimate  $z^*=v_*\tau\simeq 1\,m$.
For $\kappa(z)$ growing with $z$ faster than $z^2$, the flux is negative at $z<z^*$ and the weakly inertial distribution $z/\kappa(z)$ is normalizable at zero, while the flux is positive at  $z>z^*$ and the solution (\ref{Strong}) behaves as $n(z)\propto z^{-(m+1)/3}$, which is normalizable at infinity. In this case, one can have a steady flux state only with a source at $z\simeq z^*$; any distribution of particles with different $\tau$ undergoes dispersion with weakly inertial particles going to the minimum and strongly inertial ones going up.
On the contrary, if diffusivity grows with the distance slower than quadratic, then the fluxes must converge towards $z^*$ where all particles concentrate; such steady distribution requires an unlikely configuration of two sources and sink in between. Without sources and sinks, one can speculate that even though turbulence presents a barrier at $z^*$, particles eventually escape to infinity.

Let us now return to the quadratic case and discuss the regularization of singularities. At $z=0$, singularity of the equilibrium solution and non-integrable singularity of the flux solution with $n(z)\propto 1/z$  are both due to the assumption $\kappa(0)=0$.
Consider now $\kappa(z)=\mu (z^2+r^2)$ which is not amenable to a uniform analytic treatment, so we describe limiting cases. In the case of weak inertia, the normalized fluxless solution is $n(z)= {2Nr}/{\pi(z^2+r^2)}$ which has the limit mentioned above: $\lim_{r\to 0}n(z)=N\delta(z)$. We see that nonzero diffusivity at the minimum smears distribution whose maximum remains at the diffusivity minimum for weak inertia. The flux solution
$n(z)\propto A{z}/({z^2+r^2})$ has a flux $-A$ towards the wall. It is not normalizable at infinity i.e. implies a source at a finite distance and sink at $z=0$.
In the limit of strong inertia, steady solution with a flux directed {\it from the wall} is given by (\ref{Strong}) which is now $n(z)\propto  (z^3+3r^2z)^{-1/3} $.
Away from the minimum it has the same $1/z$ asymptotic, i.e.  it is  non-normalizable at infinity, that is can make sense only in a restricted space, for instance,  as a quasi-steady part behind a front propagating towards large $z$.
As far as the equilibrium solution is concerned, it does not exists for strongly inertial particles near a minimum with nonzero $\kappa(0)$ or an ideally reflecting wall.
The situation must change for inelastic wall, that absorbs some momentum of reflecting particles thus hampering their escape.
We hypothesize that for an intermediate case of a wall with a restitution coefficient $0\leq e\leq 1$, there must be a line of the localization-delocalization phase transition on the plane $e,I$, which starts from $e=1, I=I_c$ (reflective wall) and goes to $e\to0,
I\to\infty$ (totally inelastic); particles with $I$ above the curve escape to infinity, while those below it stay near the wall.

Another way to understand the localization-delocalization transition is to recall that a stable equilibrium distribution must decay slower than the flux solution in the direction of the flux since the flux tends to restore equilibrium \cite{ZLF,FF}. Therefore, in the low-inertia case, when the flux is negative, equilibrium state can be realized in an infinite space. In the high-inertia state, the sign is reversed and the particles go away from the minimum of diffusivity without forming any equilibrium steady state (flux state can be realized with an appropriate source). Paradoxically, even though the effect of turbophoresis is due to inertia, it reverses sign when inertial time is getting larger than the time to feel flow inhomogeneity. Of course, between two walls separated by a finite distance the fluxless solution is eventually established for any $I$.

Now let us go beyond the assumption that $\tau_c$ is the shortest time.
If $\tau\ll\tau_c$, then inertia can be neglected and particles follow the incompressible flow forming uniform concentration even for inhomogeneous $\kappa(z)$. Indeed, $n=\const$ is the only fluxless steady solution of the standard diffusion equation
$\partial_t n=\partial_z [\kappa(z)\partial_z  n]$.
The constant  distribution can take place between walls, but it is non-normalizable in the whole space, where particles escape to infinity. On the other hand, the above analysis shows that the localized steady distribution exists for $\tau\gg\tau_c$.
It is then clear that there must be a localization-delocalization transition at some critical Stokes number, $\St=\tau/\tau_c$.
To describe it analytically, let us take the double limit $\tau\to 0$, $\tau_c\to 0$, keeping $\St$ finite. In the limit, the concentration obeys the continuity equation $\partial_t n=-\partial_{z}J$, where the flux is $J=\kappa\partial_{z} n+ n\partial_{z}\kappa\, \St/(1+\St)$ (see the Appendix). Such a model is not of much practical use since generally $\St$ itself may depend on $z$ in an inhomogeneous turbulence, we use it here to illustrate the transition.
The equilibrium fluxless solution for $\kappa(z)=\mu z^m$ is $n(z)\propto z^{- {m\St}/(1+\St)}$ (a particular case of such solution for $m=4$ was obtained in \cite{Sik}). For $m=2$ the fluxless solution becomes normalizable at the critical  Stokes number, $\St_c=1$.
At exactly the same number the flux solution $n=(J/\mu z)(\St+1)/(\St-1)$ changes the  sign of the flux in agreement with the above general arguments.

In many cases, one needs to account for the gravity force. Let us show that it also can cause localization-delocalization transition in system with inhomogeneous turbulence. Consider practically important and physically interesting case of small particles in a non-uniform flow under gravity, for instance, distribution of fine dust in a turbulent wind.  Keeping $g\tau$ finite we obtain an extra term in the continuity equation: $\partial_t n=-\partial_{z}(J-g\tau n)$.
Particularly interesting is an application to the turbulent boundary layer, where $\kappa(z)=v_*z$ \cite{MY,ZA}, in this case, the dust concentration is distributed by a power law in the fluxless case:
$n(z)\propto z^{-{g\tau}/{v_*}}$.
That distribution is  normalizable  for $g\tau>v_*$, otherwise particles escape to infinity.
We thus see that if the only effect of inertia is gravitational settling, larger and heavier particles (larger $\tau$) stay close to the ground while smaller and lighter particles escape.

Yet another step towards more realistic models is to go beyond point-particle approximation. Temperature inhomogeneity in general induces a thermophoretic force proportional to the temperature gradient \emph{and the particle size}.
Taking this force into account we write the following Fokker-Planck equation for particle in the vicinity of temperature minimum
\begin{equation}
-v \partial_z \rho+\gamma\partial_{v}(v\rho)+kz\partial_{v}\rho+\gamma^2\mu z^2\partial_{v}^2\rho=0,
\end{equation}
where $k$ is called the thermophoretic mobility.
Under assumption that the coefficients $k$ and $\gamma$ are position independent we have the following probability density function for the steady state with non-zero flux $\rho(z,v)\propto z^{-2}P(v/z)$
 where $P$ is given by  (\ref{Psigma}) with ${U}(\sigma)={\gamma\sigma^2}/{2}+ {\sigma^3}/{3}+k\sigma$.
The flux can be positive or negative depending on the parameters $I$ and $W=\gamma^2/k$.
Treating thermophoretic force as an independent phenomenological parameter one can find a line of phase transition on the plane $I, W$.

The above consideration was for particle much heavier than ambient fluid whose inertia effects were neglected.
For spherical particle with density $\rho_0$ in a fluid with density $\rho$ the equation of motion has the form
$\dot v-\beta \dot u=\gamma(u-v)$, where $\beta=3\rho/(\rho+2\rho_0)$.
When $\tau,\tau_c\to0$, the equilibrium  solution for $\kappa(z)=\mu z^2$ follows the power law  $n(z)\propto z^{-2\St (1+\delta)/(1+\St)} $ with
$\delta= {\beta\St}
[(1-\beta\St)^2+\beta^2\St^2]^{-1}$,
see the Appendix. Normalizability  depends on  $\St$ and $\beta$.
Particles with different densities can thus disperse.

It is not immediately clear what are exact conditions for observing the localization-delocalization transition in real flows. We addressed here steady states or long-time regimes which in reality may be not readily observed. One must also stress that our models are gross idealizations particularly with respect to real near-wall turbulence. For example, we do not account for coherent near-wall structures which may strongly influence the particle concentration.

{\bf Acknowledgements}

The work was supported by the Minerva Foundation with funding from the German Ministry for Education and Research. We thank V. Lebedev, D. Sikovsky, E. Meiburg and K. Gawedzki for useful discussions.

{\bf Corresponding author}

Correspondence to: G. Falkovich

 {}
\vskip 0.8cm
\centerline{\bf APPENDIX} \vskip 0.5cm

\section{General relations}
\label{sec:General relations}

Consider an ensemble of inertial spherical particles embedded in random flow $\vec{u}(\vec r,t)$ of an incompressible fluid.
Every particle is assumed to be so small that the flow around it is viscous.
We neglect interaction between particles and their influence on the flow.
Coordinate $\vec r$ and velocity $\vec v$ of the particle change according to
\begin{equation}
\label{Langevin1}
\frac{d\vec v}{dt}-\beta \frac{d\vec u}{dt}=\gamma(\vec u-\vec v)+\vec{\xi}
\end{equation}
where $\gamma=1/\tau$ is the friction coefficient and $\vec{\xi}$ is a thermal Langevin force which is assumed to be delta-correlated
Gaussian white noise with zero mean, i.e., $\langle \xi_i(t)\rangle=0$, $\langle \xi_i(t_1)\xi_j(t_2)\rangle=2\gamma^2 \kappa_0\delta_{ij}\delta(t_1-t_2)$.
We  defined  $\beta=3\rho/(\rho+2\rho_0)$, where $\rho$ and $\rho_0$ are densities of the
fluid and the particle respectively \cite{Maxey_1983, Maxey_1987}.
The problem can be simplified by assuming that the particle is much denser than the ambient fluid, $\beta\ll 1$.
Neglecting the inertia of the displaced fluid we obtain
\begin{equation}
\label{Langevin2}
\frac{d\vec v}{dt}=\gamma(\vec u-\vec v)+\vec{\xi}
\end{equation}
Let us start from this equation of motion turning to general case of relation between  $\rho$ and $\rho_0$ at the end.

We treat the fluid velocity $\vec{u}(\vec r,t)$ also as a Gaussian random field, characterized
statistically by the zero mean and the pair correlation function.
The flow is assumed to be isotropic and homogeneous in time, whereas there is no homogeneity in space: typical amplitude of velocity fluctuations  depends on space coordinates.
It is appropriate to take the pair correlation of the fluid velocity in the following form
\begin{eqnarray}&&
\langle u_i(\vec r_1,t_1)u_j(\vec r_2,t_2)\rangle = h(\vec r_1)h(\vec r_2)G_{ij}(|\vec r_1-\vec r_2|/r_c)
\nonumber\\&&\times 
F(|t_1-t_2|/\tau_c)
\end{eqnarray}
where $r_c$ and $\tau_c$ are correlation length  and time, the functions $h$ and $G_{ij}$ are assumed to be smooth.
The spatially varying measure of the velocity fluctuations intensity is given by $h(\vec r)$.

Let us introduce the Stokes and Kubo numbers:
\begin{equation}
\St=1/\gamma\tau_c, \ \ \ \ \ \ \  \ \Ku=\tilde v\tau_c/r_c
\end{equation}
where $\tilde v$ is a characteristic particle velocity (which in general depends on the position since the statistics of the flow is inhomogeneous in space).
In our analysis we assume $\Ku\ll 1$, that is the particle displacement during the correlation
time $\tau_c$ is much less the correlation length $r_c$.
In this case, the fluctuations of the drag force acting on the particle are mainly produced by temporal, rather than spatial, randomness of the fluid velocity.
As a result, the momentum supplied to the particle by the drag force in a short time slot do not depend on the particle velocity.
We do not restrict Stokes number considering in details limits  $\St\ll 1$ and $\St\gg 1$ as well as the general case of relation between the particle relaxation time and the fluid velocity correlation time.


\section{Small Stokes number, $\gamma\tau_c\gg1$}
\label{sec:Small Stokes}

Let us assume the particle response time to be much smaller than the fluid velocity correlation time, $\gamma\tau_c\gg1$.
In that case one can neglect the inertia of the particle in the process of its transport by random velocity field.
Neglecting the term $\dot{\vec v}(t)$ in (\ref{Langevin2}) one finds the first-order stochastic equation
 \begin{equation}
 \label{Langeven3}
\dot{\vec r}(t)=\vec u(\vec r(t),t)+\gamma^{-1}\vec{\xi}(t)
\end{equation}
that describes the evolution on timescales $t\gg\gamma^{-1}$.
The  concentration of particles at the position $\vec r$ at the moment $t$ is
$n(\vec r, t)=\langle \delta(\vec r-\vec r(t))\rangle$,
where trajectory $\vec r(t)$ is a particular solution of the equation (\ref{Langeven3}) and the averaging is over the statistics of turbulence and thermal noise.
We now focus on the case when the characteristic time of the PDF evolution is very large compared to the fluid velocity correlation time $\tau_c$.
Then one can perform standard derivation of the Fokker-Planck equation \cite{Risken} from (\ref{Langeven3}) treating the random velocity field as a white noise
\begin{equation}
\label{kinetic1}
\partial_t n=\partial_{r_i}[\kappa_{ij}(\vec r)\partial_{r_j}n]
\end{equation}
where $\kappa_{ij}(\vec r)=\kappa_0+D_{ij}(\vec r, \vec r)$ and $D_{ij}(\vec r_1, \vec r_2)=\int_{0}^{\infty}\langle u_i(\vec r_1,t)u_j(\vec r_2,0)\rangle dt$ . 
This is the well-known advection-diffusion equation for a passive scalar in random flow \cite{Falkovich_Vergassola_2001}.
The steady state solution with zero flux is homogeneous independently on the specific form of the tensor $D_{ij}(\vec r, \vec r)$.

\section{Large Stokes number, $\gamma\tau_c\ll1$}
\label{sec:Large Stokes}

Here let us consider the opposite case when the fluid velocity correlation time is much shorter the particle response time,  $\gamma\tau_c\ll1$.
We introduce the joint probability density function (PDF) of the particle's velocity and
coordinate
\begin{equation}
\rho(\vec r, \vec v, t)=\langle \delta(\vec r-\vec r(t))\delta(\vec v-\vec v(t))\rangle
\end{equation}
where the averaging is again over the statistics of turbulence and thermal noise.
Then we can derive Fokker-Planck  equation for function ${\cal P}(\vec r, \vec v, t)$ using stochastic equation (\ref{Langevin2}) with field $\vec u(\vec r, t)$ considered as a white noise
\begin{equation}
\label{kinetic2}
\partial_t \rho=-v_i\partial_{r_i}\rho+\gamma \partial_{v_i}(v_i\rho)+\gamma^2\kappa_{ij}(\vec r)\partial_{v_i}\partial_{v_j}\rho,
\end{equation}
where $\kappa_{ij}(\vec r)=\kappa_0+D_{ij}(\vec r, \vec r)$.

We cannot solve this equation in a general case of space-dependent tensor $\kappa_{ij}(z)$.
Let us consider spatial density of the particles
\begin{equation}
n(\vec r, t)=\int \rho(\vec r, \vec v, t)d^3v
\end{equation}
Assuming that $\rho(\vec r, \vec v, t)$ decays sufficiently rapidly with $v$ and integrating the Fokker-Planck equation (\ref{kinetic2}) over velocity we find
\begin{equation}
\label{concentration1}
\partial_t n=-\partial_{r_i}\langle v_i \rangle,
\end{equation}
where here and in what follows in this Section we use a notation $\langle A\rangle=\int_{-\infty}^{+\infty}A\rho d^3v$.
The conditional moment $\langle \vec v \rangle$ is a flux (i.e. the current of the particles) at a given spatial coordinate.
In order to find the equation for the flux evolution we multiply the Fokker-Planck equation (\ref{kinetic2}) by the velocity component $v_i$ and integrate
\begin{equation}
\label{flux1}
\partial_t \langle v_i \rangle=-\partial_{r_j}\langle v_iv_j \rangle-\gamma\langle v_i \rangle,
\end{equation}
Next we need to know the quadratic moment $\langle v_iv_j \rangle$ which changes according to
\begin{equation}
\label{energy1}
\partial_t \langle v_iv_j \rangle=-\partial_{r_k}\langle v_iv_jv_k \rangle-2\gamma\langle v_iv_j \rangle+2\gamma^2\kappa_{ij}(\vec r)n,
\end{equation}
Thus we face the well-known closure problem for the hierarchy of moments: evolution equation for moment of order $m$ involves the moment of order $m+1$. The problem can be closed in the case of sufficiently small inertia when the characteristic time associated with velocity dynamics tends to zero, $\gamma^{-1}\to0$.
In this limit in the main approximation the particle feels only its local
conditions: the velocity dynamics does not have enough time to experience the variations in turbulence statistics.
Physically this means that starting from an arbitrary initial velocity distribution, a local quasi-equilibrium velocity
distribution will be established after a time of the order
 $\gamma^{-1}$ for every $\vec r$.
 After that a particle position will be undergoing a slow process of diffusion.
It should be stressed that equation (\ref{kinetic2}) is analogous to the Fokker-Plunck equation describing Brownian motion in a medium with inhomogeneous temperature \cite{Widder_1989, Stolovitzky_1998}.
Thus the technique of adiabatic elimination of the Brownian particle velocity in overdamped regime can be straightforwardly applied to our problem.

The approximate solutions of equations (\ref{flux1}) and (\ref{energy1}) on  timescales $t\gg\gamma^{-1}$ are
\begin{equation}
\langle v_i \rangle\approx-\gamma^{-1}\partial_{r_j}\langle v_iv_j \rangle
\end{equation}
\begin{equation}
\langle v_iv_j \rangle\approx \gamma\kappa_{ij}(\vec r)n,
\end{equation}
Therefore the flux can be expressed as
\begin{equation}
\langle v_i \rangle\approx -\partial_{r_j}[\kappa_{ij}(\vec r)n],
\end{equation}
that after substitution into (\ref{concentration1}) finally gives the following closed equation for concentration
\begin{equation}
\label{kinetic3}
\partial_t n=\partial_{r_i}\partial_{r_j}[\kappa_{ij}(\vec r)n].
\end{equation}

We assume that the characteristic scales of $x$- and $y$-dependence of eddy diffusivity tensor $D_{ij}(\vec r,\vec r)$ are mush larger than the inhomogeneity scale of the turbulence statistics in $z$-direction.
Then we focus on $z$-dependence and obtain a one-dimensional problem:
\begin{equation}
\label{kinetic03}
\partial_t n=\partial_{z}^2[\kappa(z)n],
\end{equation}
where $\kappa(z)=\kappa_0+D_{zz}(z)$.
The steady state with zero flux is $n(z)\propto  {1}/{\kappa(z)}$.
The stationary distribution with non-zero flux is given by
$ n(z)\propto  {z}/{\kappa(z)}$.
These solutions are zero-order terms in the expansion in powers of the inverse friction constant.


Comparing (\ref{kinetic3},\ref{kinetic03}) with (\ref{kinetic1}), one sees that the limits $\tau\to0$ and $\tau_c\to0$ do not commute.
Let us stress that  the equations (\ref{kinetic1}) and (\ref{kinetic3}) are derived from the same law of motion (\ref{Langevin1}) and describe behaviour of the particle concentration at timescales larger than $\tau_c$ and $\tau$.
The difference between these two equations is in order in which the limits $\tau\to0$ and $\tau_c\to0$ are taken.
The first choice corresponds to the small Stokes number when particle dynamics coincides with that of passive tracer.
In the opposite order of the large Stokes number, the turbulence plays a role of Brownian noise with inhomogeneous temperature.
In the next section we present a model which allows us to derive both these equations as the limiting cases of a more general diffusion equation.


\section{Inertial particle in an Ornstein-Uhlenbeck type random velocity field}

In the two preceding sections we worked with the  flow field that
is delta-correlated in time.
In this Section we go beyond this assumption proposing a simple model in which random fluctuations of the fluid velocity have  finite correlation time.
Strictly speaking, correlation time of velocity fluctuations itself may depend on coordinate in an inhomogeneous turbulence.
We, however, use here the approximation of space independent correlation time, leaving generalizations for future work.

Let us assume that the dynamics of the flow field $\vec u(\vec r,t)$ is governed  by the following first-order stochastic differential equation
\begin{equation}
\dot{\vec u}(\vec r,t)=-\frac{1}{\tau_c} \vec u(\vec r,t)+\frac{1}{\tau_c} \vec \zeta(\vec r,t)
\end{equation}
where the field $\vec{\zeta}(\vec r,t)$ is incompressible ($\diverg \vec{\zeta}(\vec r,t)=0$) and delta-correlated in time.
\begin{equation}
\label{OrnsteinUhlenbeck}
\langle \zeta_i(\vec r_1,t_1)\zeta_j(\vec r_2,t_2)\rangle = 2D_{ij}(\vec r_1,\vec r_2)\delta(t_1-t_2)
\end{equation}
Thus we model the fluid velocity as an Ornstein-Uhlenbeck process \cite{Uhlenbeck_1930, Kampen_1992}.
The correlation time of this process equals to $\tau_c$ since the pair correlation function is
\begin{equation}
\langle u_i(\vec r_1,t_1)u_j(\vec r_2,t_2)\rangle = \frac{2}{\tau_c} D_{ij}(\vec r_1,\vec r_2)e^{-|t_1-t_2|/\tau_c}
\end{equation}


Analyzing the motion of inertial particle placed into this flow, it is convenient to represent the fluid velocity field $\vec u(\vec r,t)$ and white noise $\vec \zeta(\vec r,t)$ as superposition of very large number of  spatial harmonics
\begin{equation}
\vec u(\vec r,t)=\sum\limits_{n=1}^{N}\vec u^{k_n}(t)e^{i\vec k_n\vec r}, \ \ \ \ \ \ \ \vec \zeta(\vec r,t)=\sum\limits_{n=1}^{N}\vec \zeta^{k_n}(t)e^{i\vec k_n\vec r}
\end{equation}
Then the complete system of dynamic equations for particle and fluid is
\begin{equation}
\label{system}
\left\{\begin{array}{ll}
\dot{\vec u}^{k_n}(t)=- \frac{1}{\tau_c}\vec u^{k_n}(t)+\frac{1}{\tau_c}\vec \zeta^{k_n}(t),\ \ n=1,...,N \\
\\
\dot{\vec r}(t)=\vec v(t)\\
\\
\dot{\vec v}(t)=-\gamma\vec v(t)+\gamma\vec u(\vec r(t),t)+\vec \xi(t)
\end{array} \right.
\end{equation}
where we ignore the thermal noise for the sake of brevity of the following calculations.
However the inclusion of the corresponding term into the consideration is trivial.

Thus we describe the state of the system "particle+fluid"\  by the set of $3N+6$ variables: components of the vectors $\vec r, \vec v, \vec u^{k_1}, \vec u^{k_2},...,\vec u^{k_N}$.
The joint probability density function is defined as
\begin{widetext}
\begin{equation}
{\cal P}(\vec r, \vec v, \vec u) =\langle \delta(\vec r-\vec r(t))\delta(\vec v-\vec v(t))\delta(\vec u^{k_1}-\vec u^{k_1}(t))\delta(\vec u^{k_2}-\vec u^{k_2}(t))...\delta(\vec u^{k_N}-\vec u^{k_N}(t))\rangle
\end{equation}
\end{widetext}
where $\vec r(t), \vec v(t), \vec u^{k_1}(t), ...,\vec u^{k_N}(t) $ are the particular solutions of the stochastic equations (\ref{system}) and the averaging is over the statistics of the random field $\vec{\zeta}(\vec r, t)$.
This function follows the Fokker-Planck equation
\begin{widetext}
\begin{equation}
\label{kinetic5}
\frac{\partial}{\partial t} {\cal P}=-v_i
\frac{\partial}{\partial{r_i}}{\cal P}+\gamma\frac{\partial}{\partial{v_i}}(v_i{\cal P})-\gamma\sum\limits_{n=1}^{N}e^{i\vec k_n\vec r}u_i^{k_n}\frac{\partial}{\partial{v_i}}{\cal P}
+\frac{1}{\tau_c}\frac{\partial}{\partial{u_i^{k_n}}}(u_i^{k_n}{\cal P})+\frac{1}{2\tau_c^2}\frac{\partial^2}{\partial{u_i^{k_n}}\partial{u_j^{k_m}}}[(D_{ij}^{k_nk_m}+D_{ji}^{k_mk_n}){\cal P}]
\end{equation}
\end{widetext}
where
\begin{equation}
D_{ij}^{k_nk_m}=\int e^{-i\vec k_n\vec r_1-i\vec k_m\vec r_2}D_{ij}(\vec r_1,\vec r_2)d^3r_1d^3r_2,\\
\end{equation}

The next step in our computation is a partial reduction of description by excluding the bath variables $\vec u^{k_1}, ...,\vec u^{k_N}$ and velocity $\vec v$  so that the only variables explicitly remaining is the position of the particle $\vec r$.
The spatial distribution of the particles
\begin{equation}
n(\vec r)=\int {\cal P}d^3vd^{3N}u
\end{equation}
obeys
\begin{equation}
\partial_t n=-\partial_{r_i}\langle v_i \rangle
\end{equation}
where the flux of particles is given by
\begin{equation}
\langle v_i \rangle=\int v_i{\cal P}d^3vd^{3N}u
\end{equation}
In this section the angular brackets means $\langle A \rangle=\int A{\cal P}d^3vd^{3N}u$.
Multiplying the Eq.(\ref{kinetic5}) with $v_i$ and integrating over $d^3vd^{3N}u$ one obtains the equation for evolution of the flux
\begin{equation}
\label{flux_particle}
\partial_t \langle v_i \rangle=-\partial_{r_j}\langle v_i v_j\rangle-\gamma \langle v_i \rangle+\gamma\sum\limits_{n=1}^{N}e^{i\vec k_n\vec r}\langle u_i^{k_n} \rangle
\end{equation}
The first conditional moments of the Fourier-components of the fluid velocity change in time accordingly the equation
\begin{equation}
\label{flux_fluid}
\partial_t \langle u_i^{k_n} \rangle=-\partial_{r_k}\langle u_i^{k_n} v_k\rangle-\frac{1}{\tau_c} \langle u_i^{k_n} \rangle
\end{equation}
Next we need to know quadratic conditional moments of fluid and particle velocities
\begin{eqnarray}&&
\label{quadratic1}
\partial_t \langle v_iv_j \rangle=-\partial_{r_k}\langle v_i v_jv_k\rangle-2\gamma \langle v_iv_j \rangle
\nonumber\\&&+\gamma\sum\limits_{n=1}^{N}e^{i\vec k_n\vec r}\langle u_i^{k_n}v_j \rangle
+\gamma\sum\limits_{n=1}^{N}e^{i\vec k_n\vec r}\langle u_j^{k_n}v_i \rangle
\end{eqnarray}
\begin{eqnarray}&&
\label{quadratic2}
\partial_t \langle u_i^{k_n}v_j \rangle=-\partial_{r_k}\langle u_i^{k_n} v_jv_k\rangle-\gamma \langle u_i^{k_n}v_j \rangle
\nonumber\\&& +\gamma\sum\limits_{m=1}^{N}e^{i\vec k_m\vec r}\langle u_i^{k_n}u_j^{k_m} \rangle-\frac{1}{\tau_c}\langle u_i^{k_n}v_j \rangle
\end{eqnarray}
\begin{eqnarray}&&
\label{quadratic3}
\partial_t \langle u_i^{k_n}u_j^{k_m} \rangle=-\partial_{r_k}\langle u_i^{k_n}u_j^{k_m}v_k\rangle
-\frac{2}{\tau_c}\langle u_i^{k_n}u_j^{k_m} \rangle
\nonumber\\&& 
+\frac{1}{\tau_c^2}(D_{ij}^{k_nk_m}+D_{ji}^{k_mk_n})n
\end{eqnarray}


To close this infinite set of equations for statistical moments of particle and fluid velocities, we again consider short-correlated in time fluid velocity and strong viscous damping.
Then the particle and fluid velocities evolve much faster than the concentration.
As both the particle relaxation and velocity correlation times are taken to zero the closed equation for the concentration can be derived.
We now keep $\St=\tau/\tau_c$ finite  and let $\tau\to 0$, $\tau_c\to0$.
From (\ref{quadratic1}-\ref{quadratic3}) on timescales $t\gg\tau_c,\tau$ in the main approximation we obtain
\begin{equation}
\langle v_i v_j\rangle\approx \frac{1}{2}\sum\limits_{n=1}^{N}e^{i\vec k_n\vec r}(\langle u_i^{k_n} v_j\rangle+\langle u_j^{k_n} v_i\rangle)
\end{equation}
\begin{equation}
\langle u_i^{k_n} v_j\rangle\approx \frac{\gamma}{2(1+\gamma\tau_c)}\sum\limits_{m=1}^{N}e^{i\vec k_m\vec r}(D_{ij}^{k_nk_m}+D_{ji}^{k_mk_n})n
\end{equation}
\begin{equation}
\langle u_i^{k_n}u_j^{k_m} \rangle\approx\frac{1}{\tau_c}(D_{ij}^{k_nk_m}+D_{ji}^{k_mk_n})n
\end{equation}
This gives the following expression for quadratic statistical moments of  particle velocity
\begin{equation}
\label{energy2}
\langle v_i v_j\rangle\approx\frac{\gamma}{1+\gamma\tau_c}D_{ij}(\vec r,\vec r)n
\end{equation}
Solving the equations (\ref{flux_particle}) and (\ref{flux_fluid}) on  timescales $t\gg\tau_c,\gamma^{-1}$ we have
\begin{equation}
 \langle v_i \rangle=-\frac{1}{\gamma}\partial_{r_j}\langle v_i v_j\rangle+\sum\limits_{n=1}^{N}e^{i\vec k_n\vec r}\langle u_i^{k_n} \rangle
\end{equation}
\begin{equation}
\langle u_i^{k_n} \rangle\approx-\tau_c\partial_{r_k}\langle u_i^{k_n} v_k\rangle
\end{equation}
Therefore
\begin{equation}
\partial_t n\approx \frac{1}{\gamma}\partial_{r_i}\partial_{r_j}\langle v_i v_j\rangle+\tau_c \partial_{r_i}\sum\limits_{n=1}^{N}e^{i\vec k_n\vec r}\partial_{r_j}\langle u_i^{k_n} v_j\rangle
\end{equation}
Under the condition $D_{ij}(\vec r,\vec r)=D_{ji}(\vec r,\vec r)$ we finally obtain the following closed equation for the concentration
\begin{equation}
\label{kinetic6}
\partial_t n=\frac{\St}{1+\St}\partial_{r_i}\partial_{r_j}[D_{ij}n]+\frac{1}{1+\St}\partial_{r_i}[D_{ij}\partial_{r_j}n]
\end{equation}
Note that the passive scalar advection equation (\ref{kinetic1}) is recovered in the limit as Stokes number go to zero.
The opposite limits of large Stokes number gives equation (\ref{kinetic3}).

For one dimensional problem one obtain
\begin{equation}
\label{kinetic7}
\partial_t n=\frac{\St}{1+\St}\partial_{z}^2[D_{zz}n]+\frac{1}{1+\St}\partial_{z}[D_{zz}\partial_{z}n]
\end{equation}
The equilibrium fluxless solution for $\kappa(z) = \mu z^m$ is
$n(z)\propto z^{-m\St/(1+\St)}$.
The flux solution is $n(z)\propto z^{-m+1}$.
Let us stress that these results correspond to a local equilibrium: the statistic of particle velocity at given $z$ is completely determined by the local turbulence intensity.
That approximation is valid in those regions of flow where the condition $\tilde\tau(z)\gg \gamma^{-1},\tau_c$ is satisfied.
Here we again introduce the local timescale $\tilde\tau(z)$ defined as the typical length $l$ of turbulence inhomogeneity divided by the mean local velocity $\tilde v$ that can be estimated using (\ref{energy2}).

\section{Finite density}

The above consideration was for particle which is much heavier than the ambient fluid, so that the inertia effects of the displaced fluid were neglected. For arbitrary density of the particle we return to the equation of motion (\ref{Langevin1}).
The presence of fluid acceleration means that the fluid flow can not be treated as delta-correlated in time, so we model it again by the Ornstein-Uhlenbeck process.
Substituting the expression (\ref{OrnsteinUhlenbeck}) for the acceleration of the fluid into (\ref{Langevin2}), we obtain the following system of stochastic equations for particle and fluid
\begin{equation}
\left\{\begin{array}{ll}
\dot{\vec u}(\vec r,t)=- \frac{1}{\tau_c}\vec u(\vec r,t)+\frac{1}{\tau_c} \vec \zeta(\vec r,t)\\
\\
\dot{\vec r}(t)=\vec v(t)\\
\\
\dot{\vec v}=-\gamma\vec v+(\gamma-\frac{\beta}{\tau_c})\vec u+\frac{\beta}{\tau_c}\vec \zeta
\end{array} \right.
\end{equation}

Considering again the Fokker-Planck equation for the joint probability density function of the system "particle+fluid" it is then straightforward to carry out procedure of elimination of the fast variables (particle velocity $\vec v$ and fluid velocity field $\vec u(\vec r,t)$), described in previous section.
As both the inertial and the noise characteristic times are taken to zero the equation for the concentration is as follows
\begin{eqnarray}&&
\partial_t n=\frac{(1-\beta\St)^2}{1+\St}\partial_{r_i}[D_{ij}\partial_{r_j}n]
\nonumber\\&&
+\left(\frac{\St(1-\beta\St)^2}{1+\St}+\beta^2\St^2\right) \partial_{r_i}\partial_{r_j}[D_{ij}n].\label{kinetic4}
\end{eqnarray}
Therefore for one-dimensional problem $\kappa=\mu z^m$ the concentration is distributed by a power law in the fluxless case
$
n(z)\propto z^{-m\St(1+\delta)/(1+\St)}$,
where
$
\delta= {\beta\St}[{(1-\beta\St)^2+\beta^2\St^2}]^{-1}$.

\section{Quadratic case}

For inhomogeneous turbulence no general solution of the kinetic  equation ($\ref{kinetic2}$) is available.
The only tractable case is one-dimensional motion in the vicinity of turbulence intensity minimum for which exact stationary PDF with non-zero flux can be found.
We rewrite the Fokker-Planck equation
\begin{equation}
-v \partial_z \rho+\gamma\partial_{v}(v\rho)+\gamma^2\mu z^2\partial_{v}^2\rho=0,
\end{equation}
in terms of the variable $\sigma=v/z$. Using $\rho=P(z, v/z)/z$ where $P(z, \sigma)$ is the joint PDF of $z$ and $\sigma$ we find
\begin{equation}
\mu\gamma^2\frac{\partial^2P}{\partial \sigma^2}+\gamma\frac{\partial(\sigma P)}{\partial \sigma}-\sigma z\frac{\partial P}{\partial z}+\sigma^2\frac{\partial P}{\partial \sigma}+\sigma P=0.\label{oo}
\end{equation}
Integration over $z$ produces the steady state Fokker-Planck equation
\begin{equation}
\mu\gamma^2\frac{d^2P_0}{d \sigma^2}+\gamma\frac{d(\sigma P_0)}{d \sigma}+\sigma^2\frac{d P_0}{d \sigma}+2\sigma P_0=0.\label{st}
\end{equation}
on $P(\sigma)=P(z, \sigma)dz$ with the solution
\begin{equation}
P_0\propto e^{-U(\sigma)/\mu\gamma^2}\int_{-\infty}^{\sigma}e^{U(\sigma')/\mu\gamma^2}d \sigma'\
\end{equation}
where ${U}(\sigma)={\gamma\sigma^2}/{2}+ {\sigma^3}/{3}$. There are separable solutions to Eq.~(\ref{oo}) of the form $P= {\cal Z}(z){\cal V}(\sigma)$ where
\begin{equation}
\frac{d{\cal Z}}{dz}=-\frac{\alpha}{z}{\cal Z},\ \
\label{parabolic}
\mu\gamma^2\frac{d^2{\cal V}}{d \sigma^2}+\gamma\frac{d(\sigma{\cal V})}{d \sigma}+\sigma^2\frac{d{\cal V}}{d \sigma}+(\alpha+1) \sigma{\cal V}=0,
\end{equation}
that produce one-parametric family of solutions $P_{\alpha}(z, \sigma)=z^{-\alpha}{\cal V}_{\alpha}(\sigma)$. Observing that ${\cal V}_1(\sigma)$ obeys Eq.~(\ref{st}) we find that $\rho=P_1(z, \sigma)/z$ is
\begin{equation}
\label{quadratic_flux}
\rho(z,v)\propto \frac{1}{z^2}e^{-U(v/z)/\mu\gamma^2}\int_{-\infty}^{v/z}e^{U(\sigma)/\mu\gamma^2}d \sigma\ .
\end{equation}
This solution describes the state with density $\rho(z)=\int \rho(z, v) dv\propto 1/z$. We have
\begin{eqnarray}&&
P \int v \rho(z,v) dv=P \int v {\cal V}_1(v/z) dv/z^2 \propto sign(z)\langle \sigma\rangle,\nonumber\\&&  \langle \sigma\rangle=P \int \sigma P_0(\sigma)d\sigma,
\end{eqnarray}
where the principal value is necessary due to $P_0\sim \sigma^{-2}$ at large $|\sigma|$. The proportionality constants are considered positive. We observe that for $\langle \sigma\rangle<0$ the particles converge to $z=0$, but for $\langle \sigma\rangle$ they diverge from $z=0$. Thus this solution describes the flow of particles from the sources (source) at infinity to the sink (wall) at $z=0$ if $\langle \sigma\rangle<0$. Oppositely if $\langle \sigma\rangle>0$ the solution describes the outflow of particles from the source at $z=0$.

We observe from Eq.~(\ref{parabolic}) that the large $|\sigma|$ behaviour of ${\cal V}_{\alpha}(\sigma)$ is determined by the balance of the last two terms giving ${\cal V}_{\alpha}\propto |\sigma|^{-\alpha-1}$. Integrating the equation on ${\cal V}_{\alpha}(\sigma)$ over $\sigma$ we find
\begin{equation}
\lim_{|\sigma|\to\infty}\left[\sigma^2{\cal V}_{\alpha}(|\sigma|)-\sigma^2{\cal V}_{\alpha}(-|\sigma|)\right]+(\alpha-1)P\int\limits_{-\infty}^{+\infty}\sigma{\cal V}_{\alpha} d\sigma= 0,
\end{equation}
We observe that the first term on the LHS vanishes at $\alpha\geq 1$ (when $\alpha=1$ this follows from $\lim_{|\sigma|\to\infty} {\cal V}_{\alpha}(|\sigma|)/{\cal V}_{\alpha}(-|\sigma|)=1$). When $\alpha<1$ this term is not finite. Thus the solution with finite non-zero flux is $\alpha=1$, described in the main text.

\section{Linear case}

Let us consider practically important and physically interesting case of very small inertial particles in a non-uniform flow under gravity.
In the presence of a gravitational field $\vec g$ the velocity of the particle obeys the following equation
\begin{equation}
\label{Langevin5}
\frac{d\vec v}{dt}=\gamma(\vec u-\vec v)+\vec g
\end{equation}
One can easily obtain the equation for concentration of the particles in the limit $\St\to 0$, keeping the settling velocity $g\tau$ finite
$\partial_t n=\partial_{r_i}[D_{ij}\partial_{r_j}n]+\gamma^{-1}g_i\partial_{r_i}n$.
The only difference from the equation (\ref{kinetic6}) is in presence of extra "gravity"\  term.

Let us introduce a reference system with the $z$-axis perpendicular to the ground (that is parallel to the field of gravity).
We assume that characteristic spatial scale of $x$- and $y$-dependence of tensor $D_{ij}(\vec r,\vec r)$ is mush larger than inhomogeneity scale of the turbulence statistics in vertical direction.
Then we can ignore $x$- and $y$-dependence of the concentration and write
\begin{equation}
\label{kinetic8}
\partial_t n=\partial_{z}[D_{zz}\partial_{z}n]+\gamma^{-1}g\partial_zn
\end{equation}
Particularly interesting is an application to the turbulent boundary layer, where $D_{zz}(z)=v_*z$ \cite{MY, ZA}.
The particles are distributed by a power law in the stationary fluxless case:
\begin{equation}
n(z)\propto z^{-\frac{g\tau}{v_*}}
\end{equation}
Apparently, such a distribution is realizable only when normalizable i.e. for $g\tau/v_*>1$, otherwise particles escape to infinity.
We thus see that if the only effect of inertia is gravitational settling, heavier particles (larger $\tau$) stay close to the ground while lighter particles escape.

\section{Thermophoretic force}

Yet another step towards more realistic models is to go beyond the point-particle approximation.
In the presence of a temperature inhomogeneity there will in general be a thermophoretic force proportional to
the temperature gradient and the particle size \cite{Goldhirsch}.
Taking this force into account we write the following Fokker-Planck equation for particle in vicinity of temperature minimum
\begin{equation}
-v \partial_z \rho+\gamma\partial_{v}(v\rho)+kz\partial_{v}\rho+\gamma^2\mu z^2\partial_{v}^2\rho=0,
\end{equation}
where $k$ is called the thermophoretic mobility.
This equation can be rewritten as
\begin{equation}
\mu\gamma^2\frac{\partial^2\rho}{\partial \sigma^2}+\gamma\frac{\partial(\sigma\rho)}{\partial \sigma}-\sigma z\frac{\partial\rho}{\partial z}+k\frac{\partial\rho}{\partial \sigma}+\sigma^2\frac{\partial\rho}{\partial \sigma}=0,
\end{equation}
where $\sigma=v/z$ (observe that $\rho(z, \sigma)$ differs from $P(z,\sigma)$ by a factor of $z$).

Next we separate variables $\rho= {\cal Z}(z){\cal V}(\sigma)$
\begin{equation}
\frac{d{\cal Z}}{dz}=-\frac{\alpha}{z}{\cal Z}
\end{equation}
\begin{equation}
\mu\gamma^2\frac{d^2{\cal V}}{d \sigma^2}+\gamma\frac{d(\sigma{\cal V})}{d \sigma}+k\frac{d{\cal V}}{d \sigma}+\sigma^2\frac{d{\cal V}}{d \sigma}+\alpha \sigma{\cal V}=0,
\end{equation}
Integrating last equation over velocity we find
\begin{equation}
(\alpha-2)\int\limits_{-\infty}^{+\infty}\sigma{\cal V} d\sigma\propto (\alpha-2)\int\limits_{-\infty}^{+\infty} v\rho dv= 0,
\end{equation}
Thus the state with non-zero flux corresponds to the choice $\alpha=2$ that gives the solution
\begin{equation}
\rho(z,v)\propto \frac{1}{z^2}e^{-U(v/z)/\mu\gamma^2}\int_{-\infty}^{v/z}e^{U(\sigma)/\mu\gamma^2}d \sigma
\end{equation}
with ${U}(\sigma)={\gamma\sigma^2}/{2}+ {\sigma^3}/{3}+k\sigma$.
The flux can be positive or negative depending on the parameters $I$ and $W = \gamma^2/k$.
Treating thermophoretic force as an independent phenomenological parameter one can find a line of phase
transition on the plane $I$,$W$.


\begin{thebibliography}{99}
\bibitem{Kampen}  Van Kampen, N.G. Relative stability in nonuniform temperature. \textit{IBM J. Res. Dev.} {\bf 32,} 107-111 (1988)
\bibitem{Milligen}  Van Milligen, B. P., Bons, P. D.,  Carrenras, B. A. \& Sanchez, R. On the applicability of Fick's law to diffusion in inhomogeneous systems. \textit{Eur. J. Phys.} {\bf 26,} 913 (2005)
\bibitem{Widder_1989}  Widder, M. E. \&  Titulaer, U. M. Brownian motion in a medium with inhomogeneous temperature. \textit{Physica A} {\bf 154,} 452-466 (1989).
\bibitem{Stolovitzky_1998}  Stolovitzky, G. Non-isothermal inertial Brownian motion. \textit{Phys. Lett. A}  {\bf 241,} 240-256 (1998).
\bibitem{lopez}  Lopez, C.  \&  Marconi, U. M. B. Multiple time-scale approach for a system of Brownian particles in a nonuniform temperature field. \textit{Phys. Rev. E} {\bf 75,} 021101 (2007).
\bibitem{Reeks} Reeks, M.W. Transport, mixing and agglomeration of particles in turbulent flows.
\textit{Flow, Turbulence and Combustion} {\bf 92}, 3-25 (2014)
\bibitem{Scott}  Hottovy, S., Volpe, G. \& Wehr, J. Thermophoresis of Brownian particles driven by coloured noise.
    \textit{EPL}  {\bf 99,}  60002 (2012).
\bibitem{Ci}  Caporaloni, M.,  Tampieri, F.,  Trombetti, F. \&  Vittori, O. Transfer of particles in nonisotropic air turbulence. \textit{J. Atmos. Sc.} {\bf 32,} 565-568 (1975).
\bibitem{Reek} Reeks, M.W. The transport of discrete particles in inhomogeneous turbulence. \textit{J. Aeros. Sc.} {\bf 14,} 729-739 (1983).
\bibitem{Brook} Brooke, J. W., Kontomaris, K., Hanratty, T. J. \& McLaughlin, J. B. Turbulent deposition and trapping of aerosols at a wall.\textit{ Phys. Fluids A} {\bf 4}, 825-834 (1992).
\bibitem{McL} McLaughlin, J. B.
Aerosol particle deposition in numerically simulated channel flow.
\textit{Phys. Fluids} {\bf 1}, 1211-1224 (1989).
\bibitem{Talbot}Talbot, L., Cheng, R. K., Schefer, R. W., \& Willis, D. R.
Thermophoresis of particles in a heated boundary layer.
\textit{J  Fluid Mech.} {\bf 101,} 737-758 (1980).
\bibitem{Maxey_1983}  Maxey, M.R. \&  Riley, J.J. Equation of motion for a small rigid sphere in a nonuniform flow. \textit{Phys. Fluids} {\bf 26}, 883-889 (1983).
\bibitem{Maxey_1987} Maxey, M.R. The gravitational settling of aerosol particles in homogeneous turbulence and random flow fields. \textit{J. Fluid Mech.} {\bf 174}, 441-465 (1987).
\bibitem{Risken}  Risken, H.  \textit{Fokker-Planck Equation} (Springer Berlin Heidelberg, Berlin, 1984).
\bibitem{Falkovich_Vergassola_2001}  Falkovich, G.  Gawedzki, K. \&  Vergassola, M. Particles and fields in fluid turbulence. \textit{ Rev. Mod. Phys.} {\bf 73}, 913 (2001).
\bibitem{Uhlenbeck_1930}  Uhlenbeck, G.E. \&  Ornstein, L.S. On the theory of the Brownian motion. \textit{ Phys. Rev.} {\bf 36}, 823 (1930)
\bibitem{Kampen_1992}  Van Kampen, N.G.\textit{ Stochastic Processes in Physics and Chemistry} (Elsevier, Amsterdam 1992)
\bibitem{MY} Monin, A.S. \&  Yaglom, A.M. \textit{Statistical fluid mechanics: mechanics of turbulence} (Dover, New York, 2007).
\bibitem{ZA} Zaichik, L.I,  Alipchenkov, V.M. \& Sinaiski, E.G. \textit{Particles in Turbulent Flows} (Wiley-VCH, Weinheim, 2008).
\bibitem{Goldhirsch}  Goldhirsch, I. \&  Ronis, D. Theory of thermophoresis. I. General considerations and mode-coupling analysis. \textit{Phys. Rev. A} {\bf 27,} 1616 (1983)
\bibitem{Halperin}  Halperin, B. I.  Green's functions for a particle in a one-dimensional random potential.
Phys. Rev. {\bf 139,} A104-A117 (1965)
\bibitem{FT} Feigelman, M.V. \& Tsvelik, A.M.
Hidden supersymmetry of stochastic dissipative dynamics.
\textit{Sov Phys JETP} {\bf 56}, 823 (1982)
\bibitem{Gaw2}  Gawedzki, K., 
 Herzog, D. \&  Wehr, J.
Ergodic Properties of a Model for Turbulent Dispersion of Inertial Particles.
\textit{Com. Mat. Phys.}
{\bf 308,} 49-80 (2011).
\bibitem{Deutch} Deutsch, J.M. Aggregation-disorder transition induced by fluctuating random forces. \textit{J. Phys. A} {\bf 18,} 1449-1456 (1985). 
\bibitem{WM}
 Wilkinson, M. \& Mehlig, B. Path coalescence transition and its applications.  \textit{Phys. Rev. E} {\bf 68,} 040101 (2003) .
\bibitem{MM} Martin J. \& Meiburg, E.
The accumulation and dispersion of heavy particles in forced twodimensional
mixing layers. 1. The fundamental and subharmonic cases.
\textit{Phys. Fluids} {\bf6}, 1116-1132 (1994).
\bibitem{ZLF}  Zakharov, V.E.,  L'vov, V.S. \&  Falkovich, G.E. \textit{ Kolmogorov spectra of turbulence.} (Springer, Berlin, 1991)
\bibitem{FF}  Fournier, J.D. \& Frisch, U.  d-Dimensional turbulence. \textit{Phys. Rev. A} {\bf17,} 747 (1978).
\bibitem{Sik} Sikovsky, D.P.
Singularity of inertial particle concentration in the viscous sublayer of wall bounded turbulent flows.
\textit{Flow Turb Combust.} {\bf92}, 41-64 (2014).
\end{thebibliography}
\end{document}